\documentstyle [12pt]{article}
\newcommand{\bce}{\begin{center}}
\newcommand{\ece}{\end{center}}
\newcommand{\beq}{\begin{equation}}
\newcommand{\eeq}{\end{equation}}
\newcommand{\bef}{\begin{figure}}
\newcommand{\eef}{\end{figure}}
\newcommand{\bea}{\vspace{0.25cm}\begin{eqnarray}}
\newcommand{\eea}{\end{eqnarray}}

\newcommand{\lapp}{\stackrel{<}{\sim}}
\newcommand{\gapp}{\stackrel{>}{\sim}}
\newcommand{\ba}{\begin{array}}
\newcommand{\ea}{\end{array}}

\newcommand{\ie}{{\sl i.e.~}}

\newcommand{\rhs}{{\sl rhs~}}

\newcommand{\eg}{{\sl e.g.~}}

\newcommand{\idk}{\int\frac{d^3k}{(2\pi)^3}}
\newcommand{\idp}{\int\frac{d^3p}{(2\pi)^3}}
\newcommand{\idom}{\int\frac{d\omega}{\pi}}
\newcommand{\idomp}{\int\frac{d\omega'}{\pi}}
\newcommand{\idEp}{\int\frac{dE'}{\pi}}
\newcommand{\idx}{\int\limits_{+1}^{-1}dx}

\newcommand{\singlespace}{
    \renewcommand{\baselinestretch}{1}\large\normalsize}
\newcommand{\doublespace}{
    \renewcommand{\baselinestretch}{1.6}\large\normalsize}

\newcommand{\fturbana}{\footnote[1]{also at: Department of Physics,
University of Illinois at Urbana-Champaign, 1110 West Green St.,
Urbana, IL 61801-3080, USA}}

\begin{document}
\pagestyle{empty}
\singlespace
\vspace{1.0in}
\begin{flushright}
October, 1995 \\
\end{flushright}
\vspace{1.0in}
\bce
{\large{\bf Equation of State of an Interacting Pion Gas with
Realistic $\pi$--$\pi$ Interactions}}
\vskip 1.0cm
R. Rapp and J. Wambach\fturbana

\vspace{.35in}
{\it Institut f\"ur Kernphysik (Theorie)\\
Forschungszentrum J\"ulich\\
D--52425 J\"ulich \\
Germany}
\ece
\vspace{.65in}
\begin{abstract}
\noindent
Within the finite-temperature Greens-function formalism we study
the equation of state of a hot interacting pion gas at zero
chemical potential. Employing realistic $\pi\pi$ meson-exchange
interactions we selfconsistently calculate the in-medium single-pion
selfenergy and the $\pi\pi$ scattering amplitude in the quasiparticle
approximation. These quantities are then used
to evaluate the thermodynamic potential, $\Omega_\pi$(T), from which
the state variables: pressure-, entropy- and energy-density can be
derived. In contrast to earlier calculations based on the low-energy
Weinberg Lagrangian we find an overall increase
as compared to the free-gas results. We also consider the
possibility of a dropping $\rho$-meson mass as suggested by the
'Brown-Rho Scaling' law.
\end{abstract}
\vspace{.75in}
\begin{flushright}
\singlespace
PACS Indices: 13.75.Lb\\
24.10.Cn\\  25.75.+rq
\end{flushright}
\newpage
\pagestyle{plain}
\baselineskip 16pt
\vskip 48pt

\newpage
\doublespace

\section{Introduction}
\noindent In recent years much effort has been put into the study
of ultrarelativistic heavy--ion collisions (URHIC's).
The main objective of these experiments is to create a new state of
hadronic matter, the Quark--Gluon plasma (QGP). However, it is not
settled if the current generation of experiments at the BNL--AGS and
CERN--SpS is able to generate sufficiently high energy densities
to make the transition. On the other hand, lattice gauge calculations,
although not yet at
the stage of making accurate predictions, suggest the occurrence of
a different kind of phase transition associated with the
restoration of chiral symmetry. This 'chiral phase transition'
is expected to occur prior to the deconfinement transition at a
critical temperature of $T_c^\chi= 140-170$~MeV. Precursors
of chiral symmetry restoration may well establish before that,
{\eg}the dropping of vector meson masses as proposed by Brown
and Rho~\cite{BrRo,AdBr} based on chiral effective Lagrangians and
further corroborated by Hatsuda et al.~\cite{Hats} within the
QCD sum rule approach.

In central collisions at the AGS, the midrapidity region is
characterized by highly compressed nuclear matter due to an
almost entire stopping of the Lorentz--contracted colliding
nuclei. Even at SpS energies, where the stopping power is much
less, one encounters sizable baryon densities at
central rapidity, leading to an appreciable impact on the in--medium
properties of the produced secondaries (mostly pions).
At the Brookhaven Relativistic Heavy Ion Collider (RHIC), on the
other hand, baryon densities
in the central zone are expected to be very small, such that, after
hadronization, this zone will be populated by a dense pion gas.

Many aspects of a hot, interacting pion gas have been studied in the
literature so far, {\eg} single--pion 'optical'
potentials~\cite{Shur,Sche,RW1,ChDa}, mean free paths~\cite{GoLe},
in--medium
$\pi\pi$ cross sections~\cite{ACSW,BBDS,RW1}, the equation of
state~\cite{WVPr,BuKa}, hydrodynamic properties~\cite{Gavi,Gers}
as well as numerical solutions of a bosonic Boltzmann
equation~\cite{WeBe}
simulating the dynamics of URHIC's prior to freeze out.

In this article we want to concentrate on the equilibrium
properties of a thermal pion gas, {\ie}the equation of state (EOS).
Within the relativistic virial expansion, restricted to two--body
collisions, Welke et al.~\cite{WVPr} have previously
employed an empirical
parametrization of the experimental vacuum $\pi\pi$ scattering
phase shifts in s-- and p--wave as their basic input
for calculating the
number--, energy--, pressure-- and entropy--densities
of an interacting
gas of pions. They have found that the net effect of the $\pi\pi$
interaction stems from the resonant JI=11--($\rho$--)channel leading
to an increase of the thermodynamic state variables, relative
to the free gas, for temperatures T${\gapp}$100~MeV. Quantitatively
this
contribution is comparable to what one
would expect from an admixture of free $\rho$ mesons. In other words,
the rather sharp $\rho$ resonance in the $\pi\pi$ interaction
resembles the contribution from free $\rho$ mesons, a mechanism
well--known from the Beth--Uhlenbeck formalism~\cite{BeUh}.
Bunatian and K\"ampfer~\cite{BuKa} have pursued a rather different
approach, based on finite temperature Green's functions in the
imaginary time formalism. As the $\pi\pi$ interaction they employed
the Weinberg Lagrangian~\cite{W671}, known to account for low--energy
s--wave $\pi\pi$ scattering. As a result of their selfconsistent
calculations in the Hartree approximation the thermodynamic
state variables exhibit a {\it decrease}
for temperatures T${\gapp}$ 150~MeV.

\noindent
The aim of the present paper is to recalculate the EOS by both
employing a realistic $\pi\pi$ model capable of describing the
vacuum scattering data up to rather high energies and by taking into
account the in--medium modifications of single--pions as well as
of the $\pi\pi$ scattering amplitude selfconsistently and to all
orders. Our formalism for calculating the thermodynamics will
be the one used by Bunatian and K\"ampfer, but extended beyond the
Hartree approximation. The $\pi\pi$ interaction we use  is the
meson--exchange model developed by the J\"ulich group~\cite{LoPe}
which, on a microscopic level, describes the $\pi\pi$ scattering
data in the various partial waves up to CMS energies of 1.5 GeV.

\noindent
Our article is organized as follows:
in sect.~2 we present the formalism for calculating the thermodynamic
potential $\Omega_\pi(T)$. In sect.~3 we derive from
it the thermodynamic
state variables, {\ie}the pressure--, entropy-- and energy--density,
and discuss the numerical results within the J\"ulich $\pi\pi$ model
with no in--medium modifications applied to the meson--exchange
potentials (Brueckner theory). In sect.~4 we consider the possibility
of
additional medium effects on the exchanged mesons in terms of a
dropping
$\rho$--meson mass. In sect.~5 we summarize and make some concluding
remarks.

\section{The Thermodynamic Potential of Interacting Pions}
The starting point of our analysis is the well--known expression
for the thermodynamic potential of a gas of interacting bosons
at finite temperature \cite{AGD}. In case of interacting pions it
reads
\beq
\Omega_\pi(T)=\Omega_\pi^Q(T)+\Omega_{\pi\pi}(T)
\eeq
where the first term represents the quasiparticle contribution while
the
second term arises from interactions between the quasiparticles. For
thermodynamic consistency both terms have to be considered. \\
In terms of the single--pion propagator
\beq
D_\pi(\omega_+,k)=\biggl (\omega_+^2-m_\pi^2-k^2
-\Sigma_\pi(\omega_+,k) \biggr)^{-1} \ ,
\eeq
($\omega_+\equiv \omega+i\eta$), where $\Sigma_\pi(\omega_+,k)$
denotes
the pion selfenergy, the quasiparticle contribution is given by
\beq
\Omega_\pi^Q(T)=-\frac{3}{2}\idk\int
\frac{d\omega}{\pi} \ f^\pi(\omega) \ {\rm Im} \lbrace \ln\lbrack
-D_\pi^{-1}(\omega_+,k)\rbrack-D_\pi(\omega_+,k)
\Sigma_\pi(\omega_+,k) \rbrace
\eeq
where $f^\pi(\omega)=(\exp(\omega/T) -1)^{-1}$ is the thermal
Bose factor. The interaction contribution, $\Omega_{\pi\pi}(T)$, is
the
sum of all 'skeleton diagrams' arising from a perturbative
expansion of the scattering amplitude. Within the Matsubara formalism
the contribution from $n$--th order meson exchange can be written as
\beq
\Omega_{\pi\pi}^{(n)}(T)=\frac{3}{2}\frac{1}{4n}(-T)    \idk
 \sum_{z_\nu} D_\pi(z_\nu,k) \ \Sigma_\pi^{(n)}(z_\nu,k) \ .
\eeq
This can be pictured as closing the two external legs of the pion
selfenergy.
The factor $1/4n$ corrects for overcounting~\cite{AGD} induced by
the possible ways of regenerating the $n$--th order
pion selfenergy by cutting a single--pion line in the diagrams
of $\Omega_{\pi\pi}(T)$. For the case of $\pi\pi$ contact
interactions as where considered in refs.~\cite{BuKa,BuWa} the
appropriate counting factor is 1/2$n$ due to the absence
of u--channel exchange graphs.

\noindent Let us first turn to the calculation of the two--body
interaction contribution $\Omega_{\pi\pi}(T)$.

\subsection{Lowest-Order Contributions to $\Omega_{\pi\pi}(T)$}

To lowest order ($n=1$) $\Omega_{\pi\pi}(T)$ has already been
calculated in ref.~\cite{BuKa}. In the context of a $\pi\pi$
meson--exchange potential it reads
\bea
\Omega_{\pi\pi}^{(1)}(T)=\frac{3}{8} \idk \idp  \idom  f^\pi(\omega)
Im D_\pi(\omega,k)  \times  \qquad  \qquad \qquad \qquad
 \nonumber \\
   \qquad  \idomp  f^\pi(\omega') Im D_\pi(\omega',p)
M_{\pi\pi}^{(1)}(\omega+\omega',k,p) \ ,
\eea
where $M_{\pi\pi}^{(1)}$ denotes the spin--isospin averaged,
first--order forward scattering amplitude. Employing the quasiparticle
approximation (QPA) for the single--pion propagator,
\beq
Im D_\pi(\omega,k)=-\frac{\pi}{2e_k} \lbrack \delta(\omega-e_k)-
\delta(\omega+e_k) \rbrack  \ ,
\eeq
eq.~(5) can be simplified as
\bea
\Omega_{\pi\pi}^{(1)}(T) & = & \frac{3}{8} \int \frac{k^2 dk}{2\pi^2}
\int\frac{p^2 dp}{(2\pi)^2} \idx \frac{1}{2e_k} \frac{1}{2e_p}
\times \nonumber\\
 & & \lbrace M_{\pi\pi}^{(1)}(e_p+e_k,\vec k,\vec p)
\lbrack f^\pi(e_k) f^\pi(e_p)
+(f^\pi(e_k)+1)(f^\pi(e_p)+1) \rbrack
  \nonumber\\
 & & + M_{\pi\pi}^{(1)}(e_k-e_p,\vec k,\vec p)
\lbrack f^\pi(e_k)(f^\pi(e_p)+1) + (f^\pi(e_k)+1)f^\pi(e_p) \rbrack
\rbrace  \  ,
\eea
where $x\equiv\cos\Theta \ , \Theta=\angle(\vec k, \vec p)$.
The quasiparticle energies $e_k$ are determined from the
selfconsistent solution of the Dyson equation:
\beq
e_k^2=m_\pi^2+k^2+Re\Sigma_\pi(e_k,k) \ .
\eeq
The right-hand side ({\it rhs}) of eq.~(7) diverges
due to the various '1's appearing
in the occupation factors. However, when the thermodynamic
potential is defined relative to the physical vacuum by only
considering the difference between the values at $T=0$ and finite $T$
then it remains finite. This is achieved by removing the '1's in
eq.~(7)~\cite{BuKa}. Thus
\bea
\tilde\Omega_{\pi\pi}^{(1)}(T) & \equiv & \Omega_{\pi\pi}^{(1)}(T) -
\Omega_{\pi\pi}^{(1)}(0)  \nonumber\\
& \approx & \frac{3}{8}  \int \frac{k^2 dk}{2\pi^2 2e_k}
\int\frac{p^2 dp}{(2\pi)^2 2e_p}
2 f^\pi(e_k)f^\pi(e_p) \times \nonumber\\
 & & \idx    \lbrace M_{\pi\pi}^{(1)}(e_p+e_k,\vec k,\vec p)
+ M_{\pi\pi}^{(1)}(e_k-e_p,\vec k,\vec p)\rbrace   \ .
\eea
In the following we shall suppress the tilde and always refer to
subtracted quantities. For practical evaluations, the $M$--amplitude
has to be transformed into the CMS of the two scattered pions,
thereby neglecting the medium dependence of $M_{\pi\pi}$ on the
total momentum $\vec P=\vec k +\vec p$ of the pion pair. Then
the total CMS energy is identified as
\beq
E^2\equiv s =(e_k\pm e_p)^2-(\vec k+\vec p)^2
\eeq
and, by requiring Lorentz invariance, the relative CMS
momentum of the
pions is obtained as
\beq
(\vec q)^2 =\frac{1}{s} \lbrack \frac{1}{4} (s-(e_k^2-k^2)-
(e_p^2-p^2))^2-(e_k^2-k^2)(e_p^2-p^2) \rbrack \ .
\eeq
Using $ds=-2pk dx$ we can rewrite eq.~(9) as
\beq
\Omega_{\pi\pi}^{(1)}(T)=\frac{3}{8} \int \frac{k^2 dk}{(2\pi)^2 e_k}
\int \frac{p^2 dp}{(2\pi)^2 e_p} f^\pi(e_k) f^\pi(e_p)
\lbrack I_A^{(1)}+I_B^{(1)} \rbrack
\eeq
with
\bea
I_{A/B}^{(1)} & \equiv & \int\limits_{E_{min}^{A/B}}^{E_{max}^{A/B}}
E  \ dE  \ M_{\pi\pi}^{(1)}(E,q,q) \ , \nonumber\\
E_{max}^{A/B} & = & (e_k \pm e_p)^2-k^2-p^2 + 2kp \ , \nonumber\\
E_{min}^{A/B} & = & (e_k \pm e_p)^2-k^2-p^2 - 2kp \ ,
\eea
which is the main result of this subsection.

\subsection{Higher-Order Contributions to $\bf{\Omega_{\pi\pi}(T)}$}
Within the meson--exchange framework, the full $M_{\pi\pi}$--amplitude
is obtained by solving a Lippmann--Schwinger type equation (LSE),
which, in our case~\cite{LoPe}, results from the Blankenbecler--Sugar
reduction~\cite{BBS} of the covariant Bethe--Salpeter
equation. In a given partial--wave/isospin channel $JI$
the scattering equation reads
\bea
M_{\pi\pi}^{JI}(E,q_1,q_2) & = & V_{\pi\pi}^{JI}(E,q_1,q_2)
+ \int\limits_0^{\infty} \frac{dq \ q^2}{(2\pi)^2} \
V_{\pi\pi}^{JI}(E,q_1,q)
 \   G_{\pi\pi}(E,q) \ M_{\pi\pi}^{JI}(E,q,q_2)  \ , \quad
\label{eq:Mamp}
\eea
where $G_{\pi\pi}(E,q)$ denotes the two--pion propagator of the
intermediate state and $V_{\pi\pi}\equiv M_{\pi\pi}^{(1)}$ are
the Born amplitudes (pseudopotentials) derived
from an effective meson Lagrangian. Schematically, the scattering
equation~(14) can also be written as a perturbative expansion
\bea
M & = & V  +  VGM \nonumber\\
& = & V  \quad \ +  VGV  +  VGVGV  +  \dots \nonumber\\
& \equiv & M^{(1)}  +  M^{(2)}  \ + \ \  M^{(3)}  \quad \
+  \dots \qquad .
\eea
Beyond the Born approximation, {\ie}for $n\ge 2$, the $M$--amplitude
acquires an imaginary part due to the intermediate two--pion states
being on--shell. Consequently, $\Sigma_\pi^{(n)}$ also becomes
a complex quantity. In order to evaluate eq.~(4) for
$n\ge 2$, we therefore employ the standard procedure of inserting the
spectral representations of the single--pion propagator and
the selfenergy:
\bea
D_\pi(z_\nu,k) & = & -\idomp \frac{Im D_\pi(\omega',k)}
{z_\nu-\omega'+i\eta} \nonumber\\
\Sigma_\pi^{(n)}(z_\nu,k) & = & - \idom
\frac{Im\Sigma_\pi^{(n)}(\omega,k)}{z_\nu-\omega+i\eta}
-Re \Sigma_\pi^{(n)}(0,k)  \nonumber\\
& = & -\idom Im \Sigma_\pi^{(n)}(\omega,k)
(\frac{1}{z_\nu-\omega+i\eta}
+\frac{\cal P}{\omega}) \ .
\eea
According to the findings in ref.~\cite{RW3} the dispersion relation
for the pion selfenergy has to be supplemented with a subtraction at
zero energy.

\noindent The Matsubara sum in eq.~(4) can now be performed
analytically. The imaginary part of the pion selfenergy has been
derived
in ref.~\cite{RW3} and we can immediately generalize the
result to the
$n$--th order contribution:
\beq
Im \Sigma_\pi^{(n)}(\omega_+,k)=-\int\frac{d^3p}{(2\pi)^3}
\int\frac{d\omega '}{\pi} Im M_{\pi\pi}^{(n)}(\omega_++\omega ',
\vec k, \vec p) Im D_\pi(\omega_+',p)
\lbrack f^\pi(\omega ')-f^\pi(\omega+\omega ')\rbrack \ .
\eeq
Injecting eqs.~(16),~(17) into eq.~(4) results in the following,
exact expression for the $n$--th
order scattering contribution to the thermodynamic potential:
\bea
\Omega_{\pi\pi}^{(n)}(T) & = & -\frac{3}{2}\frac{1}{4n}\idk\idp
\idEp \idom \idomp Im D_\pi(\omega',p) Im D_\pi(\omega,k) \times
\qquad
\nonumber\\
 & & Im M_{\pi\pi}^{(n)}(E',\vec k,\vec p) \lbrace
\frac{\lbrack f^\pi(\omega')-f^\pi(E')\rbrack \lbrack \omega
f^\pi(\omega)-(E'-\omega')f^\pi(E'-\omega')\rbrack}
{(\omega_++\omega'-E')(E'-\omega')} \rbrace \ .
\eea
Note that the imaginary part of the \rhs, given by the
$\delta$--function part of $ 1/(\omega_++\omega'-E')$, vanishes
due to the simultaneous disappearance of the second occupation factor
-- as it should, since $\Omega_{\pi\pi}(T)$ is a real quantity.
Applying the QPA, eq.~(6), and performing the trivial angular
integrations leads to
\bea
\Omega_{\pi\pi}^{(n)}(T) & = & -\frac{3}{2}\frac{1}{4n}
\int \frac{k^2 dk}
{2\pi^2 2e_k} \int\frac{p^2dp}{(2\pi)^2 2 e_p} \idx  \
\int\limits_0^\infty \frac{dE'}{\pi} Im M_{\pi\pi}^{(n)}(E',q,q)
\times
\nonumber\\
 & & 2 \lbrack F^\pi(E',k,p)+G^\pi(E',k,p)\rbrack \ ,
\eea
where
\bea
F^\pi(E',k,p) & = &  \quad \frac{\lbrack f^\pi(e_p)-f^\pi(E')\rbrack
\lbrack
e_k f^\pi(e_k)-(E'-e_p)f^\pi(E'-e_p)\rbrack}{(e_k+e_p-E')(E'-e_p)}
\nonumber\\
 & & +\frac{\lbrack f^\pi(e_p)+f^\pi(E')\rbrack \lbrack
e_k f^\pi(e_k)-(E'+e_p)f^\pi(E'+e_p)\rbrack}{(e_k+e_p+E')(E'+e_p)}
\nonumber\\
G^\pi(E',k,p) & = & - \frac{\lbrack f^\pi(e_p)-f^\pi(E')\rbrack
\lbrack
e_k f^\pi(e_k)-(E'-e_p)f^\pi(E'-e_p)\rbrack}{(-e_k+e_p-E')(E'-e_p)}
\nonumber\\
 & & +\frac{\lbrack f^\pi(e_p)+f^\pi(E')\rbrack \lbrack
e_k f^\pi(e_k)-(E'+e_p)f^\pi(E'+e_p)\rbrack}{(e_k-e_p-E')(E'+e_p)} \ .
\eea
The second term in $F^\pi$ arises from the negative-energy part
of the $E'$--integration. It is obtained from the first term when
replacing
$E'\rightarrow (-E')$ and introducing an overall minus sign due to the
symmetry $Im M_{\pi\pi}(-E'+i\eta)=-Im M_{\pi\pi}(E'+i\eta)$.
Similarly, the
function $G^\pi$ is generated from $F^\pi$ when replacing
$e_k\rightarrow (-e_k)$
with an overall minus sign due to the antisymmetry of the single--pion
spectral function, eq.~(6).
Since the 4--momenta
$(\omega, \vec k) , \ (\omega',\vec p)$ enter eq.~(18) on equal
footing
we obtain an additional factor of 2 for the negative energy
contributions of $Im D_\pi(\omega',p)$.      \\
It is not to be expected that, evaluating
$\Omega_{\pi\pi}^{(n)}(T)$ order by order, leads to a convergent
series.
For the LSE~(15) the Born series does not converge and, although
there appears an additional factor 1/$n$ when summing the
thermodynamic potential
\beq
\Omega_{\pi\pi}(T)=\sum_{n=1}^\infty \Omega_{\pi\pi}^{(n)}(T) \ ,
\eeq
it may not change the convergence behavior of the series.
In particular,
the s--channel pole graphs ({\ie}genuine $\pi\pi$--resonances)
turn out
to be problematic, since they exhibit non--integrable  singularities
in each order $n\ge 2$. In the following we will discuss them in
more detail.
\subsubsection{s--Channel Pole graphs}
In the J\"ulich model of Lohse et al.~\cite{LoPe} a genuine resonance
is
characterized by a separable Born term of the form
\beq
V_\alpha^{JI}(E',q_1,q_2)=v_{\pi\pi\alpha}(q_1) D_\alpha^0(E')
v_{\pi\pi\alpha}(q_2)
\eeq
with the bare resonance propagator
\beq
D_\alpha^0(E')=\frac{1}{E'^2-(m_\alpha^{(0)})^2} \ ,
\eeq
$m_\alpha^{(0)}$ being the bare mass of the resonance $\alpha$ with
spin--isospin $JI$. The vertex functions contain coupling constants,
isospin and form factors as well as kinematic factors specific to
the spin--momentum part of the coupling.

\noindent As is well known, the LSE can be solved analytically
for the case of a separable potential~\cite{JoLe} which yields
\bea
M_\alpha^{JI}(E',q_1,q_2) & = & v_{\pi\pi\alpha}(q_1) D_\alpha^0(E')
v_{\pi\pi\alpha}(q_2) \sum_{n=1}^\infty \lbrack \frac{I_\alpha(E')}
{E'^2-(m_\alpha^{(0)})^2} \rbrack^{n-1}  \nonumber\\
& = & \frac{v_{\pi\pi\alpha}(q_1) \ v_{\pi\pi\alpha}(q_2)}{E'^2-
(m_\alpha^{(0)})^2-I_\alpha(E')} \ ,
\eea
where
\beq
I_\alpha(E')=\int q^2 dq  \ v_{\pi\pi\alpha}(q)^2 \ G_{2\pi}(E',q) \
\eeq
is just the intermediate $\pi\pi$ bubble. In eq.~(24) the summation
index is chosen such that $n=1$ corresponds to the
lowest--order (Born)
term. Apparently the $n$--th order contribution in eq.~(24) exhibits
a pole of order $n$ at $E'=m_\alpha^{(0)}$, leading to
divergencies in
the $E'$--integration of eq.~(19). The infinite sum of $\pi\pi$
bubbles,
on the other hand, simply renormalizes the bare propagator
generating a mass shift and a finite width, given by
$Re I_\alpha(E')$
and $Im I_\alpha(E')$, respectively. As we shall show now, this
implies
the possibility of summing the $\Omega_{\pi\pi\alpha}^{(n)}(T)$
contributions to all orders, thereby avoiding any divergencies.
The evaluation of eq.~(21) for pure resonance scattering involves the
following sum:
\bea
\Omega_{\pi\pi\alpha}(T) & \propto & \sum_{n=2}^{\infty} \frac{1}{n}
Im M_{\pi\pi\alpha}^{(n)}(E',q,q) \nonumber\\
& = & v_{\pi\pi\alpha}(q)^2 D_\alpha^0(E') Im \lbrack
\sum_{n=2}^\infty \frac{1}{n} (\frac{I_\alpha(E')}{(E'^2-
(m_\alpha^{(0)})^2})^n \rbrack  \nonumber\\
& = & v_{\pi\pi\alpha}(q)^2  Im \lbrack \frac{1}{I_\alpha(E')}
\sum_{n=1}^\infty \frac{1}{n} (\frac{I_\alpha(E')}{(E'^2-
(m_\alpha^{(0)})^2})^n \rbrack   \  ,
\eea
where we have added a zero in form of the $(n=1)$--term (Born term),
which
does not generate any imaginary part. Using the identity
\beq
\ln (1-x)=-\sum_{n=1}^\infty \frac{x^n}{n}  \  ,
\eeq
we are now able to perform the infinite sum in eq.~(26). The resulting
thermodynamic potential reads
\bea
\Omega_{\pi\pi\alpha}(T) & = & -\frac{3}{8}\int \frac{k^2 dk}
{2\pi^2 2e_k} \int\frac{p^2dp}{(2\pi)^2 2 e_p} \idx
\int\limits_0^\infty \frac{dE'}{\pi} SI(\alpha) v_{\pi\pi\alpha}(q)^2
\times \nonumber\\  & &
Im \lbrace \frac{(-1)}{I_\alpha(E')} \ln \lbrack 1-\frac{I_\alpha(E')}
{E'^2-(m_\alpha^{(0)})^2} \rbrace  \ 2 \
\lbrack F^\pi(E',k,p)+G^\pi(E',k,p)\rbrack \ ,
\eea
with
\beq
SI(\alpha)=\frac{(2J+1)(2I+1)}{3}
\eeq
being the spin--isospin weighting factor of the corresponding $\pi\pi$
resonance channel. When numerically evaluating eq.~(28) in the
J\"ulich Model,
one obtains results far too large to be realistic.  A closer
inspection
shows that, for fixed momentum $k$, the $p$--integration dominantly
picks up
high--momentum components. The reason for that can be traced back
to the
weak suppression in the (dipole) form factors entering the vertex
functions
$v_{\pi\pi\alpha}(q)$,
\beq
F(q)=(\frac{2\Lambda_\alpha^2+m_\alpha^2}{2\Lambda_\alpha^2
+4\omega_q^2})^2 \ .
\eeq
In the limit of large p--values, $p\gg k$, the CMS momentum $q$
behaves like $q^2 \propto p$, {\ie}one essentially looses four  powers
of momentum in $v_{\pi\pi\alpha}(q)^2\propto F(q)^2$. Note that the
occupation factors $F^\pi$ and $G^\pi$, eq.~(20), contain terms
independent of $f^\pi(e_p)$ such that no 'thermal' suppression is at
work. To overcome this unrealistic behavior we have decided to
replace the
form factor at each vertex by
\beq
F(q) \rightarrow \sqrt{F(k) F(p)} =
(\frac{2\Lambda_\alpha^2+m_\alpha^2}
{2\Lambda_\alpha^2+4\omega_k^2}) (\frac{2\Lambda_\alpha^2+m_\alpha^2}
{2\Lambda_\alpha^2+4\omega_p^2})  \  ,
\eeq
which restricts the relevant momentum ranges in eq.~(28) to a
realistic domain of $p,k \lapp \Lambda_\alpha$ ({\eg}$\Lambda_\rho=
3.3$~GeV in the J\"ulich model).
\subsubsection{t--Channel Vector-Meson Exchange}
In the J\"ulich model the dominant contribution to the
low--energy $\pi\pi$
interaction is generated by t--channel exchange
of the $\rho$ meson. The corresponding Born amplitude is of the
non--separable form
\beq
V_{\rho ex}^{JI}(E',q_1,q_2)=v_{\pi\pi\rho ex}(E',q_1,q_2)
 \ D_\rho(E',q_1,q_2) \ ,
\eeq
where
\beq
D_\rho(E',q_1,q_2)=\frac{1}{t-m_\rho^2}
\eeq
is the effective $\rho$ t--channel propagator with 'renormalized'
(physical) mass, $m_\rho$=770~MeV, and 4--momentum transfer
$t=(q_2-q_1)^2$.  The non--separability evades both an
analytical solution of the scattering eq.~(14) and a closed summation
of the thermodynamic potential eq.~(21). Thus we are forced to
evaluate
the t--channel exchange contributions to $\Omega_{\pi\pi}(T)$ order
by order. Fortunately there are two circumstances that will allow us
to get a realistic estimate:
\begin{itemize}
\item[--] the $\rho$ t--channel exchange propagator, eq.~(33),
does not
exhibit any poles in the kinematical region relevant for calculating
the $n$--th order contribution to $\Omega_{\pi\pi}(T)$,
thus ensuring the absence of divergencies order by order;
\item[--] as it will turn out from the numerical results in sect.~3.3
the series eq.~(21) rapidly converges when summing the closed
ladder sum ({\eg}the third order contribution is typically 10\%
of the second order contribution).
\end{itemize}
When evaluating eq.~(19) by using the expansion in eq.~(15) with the
pseudopotential~(32) we encounter the same problem as was pointed
out at the end of sect.~2.2.1, namely unrealistically
large contributions from very high momenta. Therefore, as in
the case of pole graphs, we replace the form factor entering the
vertex
functions in eq.~(32) by
\beq
F(q,q)^2 \rightarrow F(k) F(p) \ .
\eeq
By a slight readjustment of the form factor we made sure that
the exact $n$-th order $\rho$ exchange amplitude is reasonably well
reproduced when using this separable approximation.

\noindent
Another characteristic feature of the J\"ulich model is the
occurrence of a scalar--isoscalar bound state in the coupled $K\bar K$
channel, just below the $K\bar K$ threshold. This bound state is
generated by a strong attraction between kaon and antikaon, mediated
by t--channel exchange of the vector mesons $\rho (770)$,
$\omega (782)$ and $\phi (1020)$. It leads to the characteristic sharp
rise in the $\pi\pi \rightarrow \pi\pi$ phase shifts
$\delta^{00}$ in vicinity of the $K\bar K$ threshold and is
interpreted as the $f_0(980)$.
To incorporate this resonance--like feature of the $\pi\pi$ phase
shifts in our calculation of $\Omega_{\pi\pi}(T)$ we proceed as
follows: rather than summing the ladder diagrams involving the
$K\bar K$ intermediate states and their direct interaction we
simulate the $f_0(980)$ as a genuine resonance, {\ie}we choose the
resonance parameters such that the full $\delta^{00}_{\pi\pi}$ phase
shifts can be described without any coupling to the $K\bar K$ channel.
The fully iterated $f_0(980)$--polegraph contribution to
$\Omega_{\pi\pi}(T)$ is then calculated as outlined in sect.~2.2.1
(eq.~(28)).

Before ending this section  we should  point out that we have
neglected any interference terms between the s--channel pole graphs
('bubble' sum) and the t--channel exchange graphs ('ladder' sum).
The following two arguments make us believe that this is a reasonable
approximation:
\begin{itemize}
\item[--] the $\pi\pi$ interaction in the resonant channels
({\ie}the $JI$=11 channel with the $\rho (770)$--resonance and the
$JI$=20 channel with the $f_2(1270)$--resonance) is largely driven
by polegraph contributions such that the additional incorporation of
t--channel exchange processes is expected to have a
quantitatively small
effect; in the case of the $\rho$-- and $f_2$--channels the
$\rho$ t--channel exchange is attractive and leads to a rather
small change of the resonance when evaluated with  the
combined ladder--/bubble--sum. Our results for the polegraph
contribution, $\Omega_{\pi\pi\alpha}'(T)$, should therefore be
considered as a lower estimate of the exact contribution; the
situation is less clear for the (simulated) $f_0(980)$ in the
$JI$=00 channel, but as it will turn out from the numerical
results the $f_0$--bubble sum gives a very small contribution
by itself;
\item[--] as already mentioned in sect.~2.2.2 the $\rho$ t--channel
exchange contribution to $\Omega_{\pi\pi}(T)$ converges rapidly
with the ($n$+1)--th order being typically down by
10\% from the $n$--th order.
\end{itemize}
We now turn to the evaluation of the single--(quasi--)pion part,
$\Omega_\pi^Q(T)$, of the thermodynamic potential.
\subsection{Quasiparticle Contributions: $\Omega_{\pi}^Q(T)$}
In evaluating the contributions to $\Omega_{\pi}^Q(T)$ we shall also
apply the QPA to the medium--modified single--pion spectrum, given by
eq.~(2). We essentially follow the steps outlined in ref.~\cite{CaPe}.
\\
First we make use of the analytic properties of the one--pion
propagator to transform the negative energy part of the
$\omega$--integration to positive values, namely
\beq
D_\pi(\omega_+,k)=D_\pi^*(\omega_-,k) \ .
\eeq
Using the identity
\beq
Im[\ln(z)]=-Im[\ln(z^*)]
\eeq
for any complex number $z$, and removing the infinite vacuum part (as
discussed in sect.~2.1) results in
\beq
\Omega_\pi^Q(T)=-\frac{3}{2}\idk\int\limits_0^\infty
\frac{d\omega}{\pi} \ 2 f^\pi(\omega) \ Im \lbrace \ln\lbrack
-D_\pi^{-1}(\omega_+,k)\rbrack-D_\pi(\omega_+,k)
\Sigma_\pi(\omega_+,k) \rbrace \ .
\eeq
The first term in the braces can be rewritten by using the identity
\beq
Im\lbrack \ln (-D_\pi^{-1}(\omega,k))\rbrack =\pi \Theta \lbrack
Re D_\pi^{-1}(\omega,k)\rbrack - \arctan\lbrack \frac{-Im \Sigma_\pi
(\omega,k)}{Re D_\pi^{-1}(\omega,k)}\rbrack \ .
\eeq
Neglecting the imaginary part of $\Sigma_\pi$ (which we will
call scheme 'a') eq.~(37) can be simplified as
\beq
\Omega_\pi^{Q,a}(T)=\frac{3}{2\pi^2} \int dk k^2 \lbrace T
\ln (1-e^{-e_k/T})-\frac{1}{2e_k} f^\pi(e_k) Re \Sigma_\pi(e_k,k)
\rbrace      \ .
\eeq
The first term corresponds to the contribution of 'free'
quasiparticles
with modified dispersion relation $e_k=e(k)$, whereas the second
term arises from an additional, dynamically generated, mean field.
Their qualitative behavior is in line with the naive expectation
that the pressure, $p=-\Omega$, is reduced for heavier particles
but increases for a repulsive mean field and vice versa. \\
One can go one step further by including the imaginary part of the
pion selfenergy in eq.~(37) (which we will call scheme 'b'). Then
$\Omega_\pi^Q(T)$ receives further contributions and reads
\bea
\Omega_\pi^{Q,b}(T) & = & \frac{3}{2\pi^2} \int dk k^2 \lbrace T
\ln (1-e^{-e_k/T})- \int\limits_0^\infty \frac{d\omega}{\pi}
f^\pi(\omega) \arctan\lbrack \frac{Im \Sigma_\pi(e_k,k)}
{\omega^2-\omega_k^2- Re \Sigma_\pi(e_k,k)} \rbrack  \rbrace
\nonumber\\
 & & + \frac{3}{2\pi^2} \int dk k^2 \int\limits_0^\infty
\frac{d\omega}{\pi} f^\pi(\omega) \frac{Im\Sigma_\pi(e_k,k)
(\omega^2-\omega_k^2)}{(\omega^2-e_k^2)^2+(Im\Sigma_\pi(e_k,k))^2}
  \ .
\eea
We shall consider both approximations, eqs.~(39) and (40), in our
numerical calculations discussed in the next chapter.
\section{Thermodynamic State Variables and Numerical Results}
\subsection{Pressure, Entropy and Energy}
Having the thermodynamic potential at hand, the state variables:
pressure--density, $p(T)$, entropy--density, $s(T)$,
and energy--density, $\epsilon(T)$, can be calculated by means
of the standard thermodynamic relations at zero chemical
potential:
\bea
p(T) & = & -\Omega(T) \\
s(T) & = & -\frac{\partial \Omega(T)}{\partial T}  \\
\epsilon(T) & = & T s(T)-p(T)  \ .
\eea
As elaborated in refs.~\cite{CaPe,AGD} the partial
derivative w.r.t.~$T$
in eq.~(42) only acts on the explicit temperature dependence of the
occupation factors due to the stationarity condition~\cite{CaPe,AGD}
\beq
\frac{\delta \Omega}{\delta \Sigma}=0 \ .
\eeq
Thus we obtain for the quasiparticle contributions to the entropy
from eq.~(39):
\bea
s_\pi^{Q,a}(T) & = & \frac{3}{2\pi^2} \int dk k^2
\lbrace (1+f^\pi(e_k))
\ln (1+f^\pi(e_k))-f^\pi(e_k) \ln(f^\pi(e_k))  \nonumber\\
 & & \qquad\qquad\quad -\frac{1}{2e_ k} \frac{\partial f^\pi(e_k)}
{\partial T}  Re \Sigma_\pi(e_k,k) \rbrace      \ ,
\eea
or, when including a finite width for the pions, from eq.~(40):
\bea
s_\pi^{Q,b}(T) & = & \frac{3}{2\pi^2} \int dk k^2  \lbrace
(1+f^\pi(e_k)) \ln (1+f^\pi(e_k))-f^\pi(e_k) \ln(f^\pi(e_k))
\nonumber\\
  & & \qquad \qquad \quad
-\int\limits_0^\infty \frac{d\omega}{\pi}
\frac{\partial f^\pi(\omega)}{\partial T}
\arctan\lbrack \frac{Im \Sigma_\pi(e_k,k)}
{\omega^2-\omega_k^2- Re \Sigma_\pi(e_k,k)} \rbrack
\nonumber\\
 &  & \qquad \qquad \quad
+ \int\limits_0^\infty
\frac{d\omega}{\pi} \frac{\partial f^\pi(\omega)}{\partial T}
\frac{Im\Sigma_\pi(e_k,k) (\omega^2-\omega_k^2)}
{(\omega^2-e_k^2)^2+(Im\Sigma_\pi(e_k,k))^2} \rbrace
  \ ,
\eea
which is consistent with the expressions quoted in ref.~\cite{CaPe}.
In our framework of meson-exchange interactions, the skeleton
contributions to the entropy (denoted by $-\partial\Phi / \partial T$
in ref.~\cite{CaPe}) arise in lowest order from eq.~(12),
\beq
s_{\pi\pi}^{(1)}(T)=-\frac{3}{8} \int \frac{k dk}{(2\pi)^2e_k}
\int \frac{p dp}{(2\pi)^2e_p} [I_A^{(1)}+I_B^{(1)}] [\frac{\partial
f^\pi(e_k)}{\partial T} f^\pi(e_p)+f^\pi(e_k)
\frac{\partial f^\pi(e_p)}
{\partial T} ] \ ,
\eeq
for the t--channel $\rho$ exchange graphs from eq.~(19),
\bea
s_{\pi\pi}^{(n)}(T) & = & -\frac{3}{2}\frac{1}{4n}
\int \frac{k^2 dk}
{2\pi^2 2e_k} \int\frac{p^2dp}{(2\pi)^2 2 e_p} \idx  \
\int\limits_0^\infty \frac{dE'}{\pi} Im M_{\pi\pi}^{(n)}(E',q,q)
\times
\nonumber\\
 & & 2 \lbrack \frac{\partial F^\pi(E',k,p)}{\partial T}
 +\frac{\partial G^\pi(E',k,p)}{\partial T} \rbrack \ ,
\eea
and an analogous expression for the s--channel pole graphs
from eq.~(28) for $s_{\pi\pi\alpha}(T)$. \\
To check our results with respect to thermodynamic consistency,
we will also consider an alternative way of evaluating the
pressure density, namely
\beq
p_\pi(T)=\int\limits_0^T dT' \ s_\pi(T') \ .
\eeq

\subsection{Selfconsistent Brueckner Scheme}
As pointed out in sect.~2.1 the in--medium pion dispersion relation
(or pion selfenergy) has to be determined selfconsistently from
the Dyson eq.~(8). In the QPA the pion selfenergy is given in terms
of the in--medium and on--shell $\pi\pi$ forward scattering
amplitude as~\cite{RW1}
\beq
\Sigma_\pi(e_k,k)=\frac{1}{k}\int\limits_{0}^{\infty}
\frac{dp}{(2\pi)^2}
\frac{p}{2e_p}  [f^\pi(e_p)-f^\pi(e_k+e_p)]
\int\limits_{E_{min}}^{E_{max}} dE_{cms} E_{cms}
M_{\pi\pi}(E_{cms}) \ .
\eeq
Here we have neglected the numerically very small
contributions~\cite{RW3} from thermal excitations
(as is usually done in the literature).  The
in--medium scattering amplitude, to be calculated from eq.~(15),
depends again on the pion selfenergy $\Sigma_\pi$ through the
in--medium two--pion propagator of the intermediate state:
\beq
G_{2\pi}(E,q)=\frac{1}{e_q}\frac{1+2f^\pi(e_q)}{E^2-4(\omega_q^2+
\Sigma_\pi(e_q,q))} \ .
\eeq
Thus eqs.~(8),~(14),~(50),~(51) define a selfconsistency problem
of Brueckner type which we solve (at fixed temperature) by
numerical iteration as discussed in ref.~\cite{RW1}. \\
The converged results for the pion selfenergy and the in--medium
$M_{\pi\pi}$--amplitude are then used to calculate
the thermodynamic potential and state variables as described in
sect.~2 and in sect.~3.1, respectively.
According to scheme 'a' or 'b' (see sect.~2.3) we, respectively,
either neglect or include $Im\Sigma_\pi$ when evaluating the
two--pion propagator eq.~(51).
\subsection{Numerical Results and Discussion}
For given value of the temperature we selfconsistently calculate
the in--medium $\pi\pi$ amplitude as well as the pion
selfenergy, which are
then further used to evaluate the various contributions
to the thermodynamic
potential, namely
\begin{itemize}
\item[(i)] the lowest--order skeleton graphs, eq.~(12);
\item[(ii)] the resonance--$\pi\pi$ bubble skeleton graphs to all
orders, eq.~(28), for the resonances $\alpha=\rho(770), f_2(1270)$ and
the simulated $f_0(980)$;
\item[(iii)] the second-- and third--order skeleton graphs for $\rho$
t--channel exchange, eq.~(19);
\item[(iv)] the single--(quasi--)pion contributions given by
eq.~(39) or~(40).
\end{itemize}
Let us first concentrate on scheme 'a' in which we neglect any
imaginary part of the pion selfenergy. The full results for
pressure--, entropy-- and energy--density (full lines)
are contrasted in fig.~1 with the results for free pions
(dashed--dotted lines) and free $\rho$ mesons
(dotted lines). First we note that the quasiparticle
contribution eq.~(39) coincides at all temperatures within less
than 1\% with the EOS of free pions,
\beq
\Omega_\pi^{free}(T)=\frac{3T}{2\pi^2} \int dk k^2
\ln(1-e^{-\omega_k/T}) \ .
\eeq
The reason for that is an almost exact cancellation of the mean field
contribution (second term in eq.~(39)) and the kinematic modification
caused by the replacement $\omega_k\rightarrow e_k$ in the
exponential of the logarithm in the first term of eq.~(39). But even
the absolute magnitudes of these medium--induced single--particle
modifications do not exceed 3\% at any temperature. These findings
are at variance with the results of ref.~\cite{BuKa}, where a general
{\it decrease} of the thermodynamic state variables compared to the
free pion gas was found. This decrease is due to an increase of the
in--medium pion mass dominantly leading to a suppression of the
ln--term in eq.~(39). The increase of the pion mass stems from the
net repulsion of the low--energy s--wave $\pi\pi$ interaction
(Weinberg Lagrangian). In our analysis, however, it turns out that
it is important to include both a higher energy range of the
$\pi\pi$ interaction (essentially up to 1 GeV two--pion CMS energy)
and higher partial waves. {\sl E.g.} the resonant p--wave channel
significantly contributes to $Re \Sigma_\pi(e_k,k)$, whereas there
are large cancellations in both s-- and d--wave between the
attractive $I=0$-- and the repulsive $I=2$--channels~\cite{WVPr,Rdth}.

\noindent Let us now come to the discussion of the skeleton diagrams.
The effect of the first order graphs, eq.~(12), is small: their
combined contribution does not exceed 1\% of $\Omega_\pi^{free}(T)$
in the considered temperature range. The most important role is
played by the s--channel $\rho$--pole graphs, see {\eg}table 1.
It is remarkable that nearly half of the contribution to
$p_{\pi\pi\rho}(T)$, $s_{\pi\pi\rho}(T)$ and
$\epsilon_{\pi\pi\rho}(T)$ is due to negative energy contributions
represented by the $G^\pi$--term in eq.~(28). Without it the
values for $p_{\pi\pi\rho}(T)$, $s_{\pi\pi\rho}(T)$ and
$\epsilon_{\pi\pi\rho}(T)$ lie slightly below the free $\rho$--gas
values $p_{\rho}^{free}(T)$, $s_{\rho}^{free}(T)$ and
$\epsilon_{\rho}^{free}(T)$ (dashed--double--dotted lines in fig.~1).
This is in agreement with the findings of ref.~\cite{WVPr}, {\ie}that
the $\pi\pi$ p--wave interaction effect resembles the presence of free
$\rho$--mesons, the small suppression compared to the free $\rho$--gas
values in our case being caused by the finite width of the
$\rho$--resonance (which is even further broadened in medium) and the
missing t--channel exchange diagrams. The latter, when implemented
in the polegraph -- $\pi\pi$ bubble sum, would lead to a slightly
stronger renormalization of the bare $\rho$. \\
The effect of other $\pi\pi$ resonances is small:
$p_{\pi\pi f_2}(T)= 3.8\% p_{\pi}^{free}$,
$p_{\pi\pi f_0}(T)= 0.3\% p_{\pi}^{free}$ at T=200~MeV, and similarly
for entropy-- and energy--density.
A somewhat larger contribution is generated by skeleton diagrams
from second order $\rho$ t--channel exchange, see {\eg}table 2.
The third order values are already one order of magnitude smaller. \\
The overall picture emerging for the temperature dependence of the
thermodynamic state variables is quite similar for all three:
we find an appreciable enhancement of the selfconsistently
calculated pressure, entropy and energy for the interacting gas
compared to free pions for temperatures $T\ge 100MeV$, which even
exceeds the values for a free gas of pions and $\rho$--mesons. This
excess can be traced back to negative energy contributions,
naturally arising within our formalism, and low--energy t--channel
exchange of virtual $\rho$--mesons between two pions. At very high
temperatures ($T>200MeV$) the thermal broadening of the
$\pi\pi$ resonances, in particular the dominantly contributing
$\rho$(770), becomes so large that they no longer resemble the
effect of an independent particle species (remember from the
Beth--Uhlenbeck formalism that a sufficiently sharp two--particle
resonance thermodynamically acts like an independent species
corresponding to the quantum numbers of the resonance). \\
When including a finite width, $Im \Sigma_\pi$,
for the quasi--pions ($\equiv$ scheme 'b',
as described in sect.~3.2.) we observe a slight
overall enhancement of $p_\pi(T)$, $s_\pi(T)$ and $\epsilon_\pi(T)$
(dashed lines in fig.~1) compared to the results when
neglecting $Im \Sigma_\pi$. The main source for this increase
are the second order $\rho$ t--channel skeleton graphs, the $\rho$
pole skeleton graphs and the quasiparticle contributions eq.~(40),
compare table 3.
In case of the skeleton diagrams this increase is simply due to the
fact that $Im M_{\pi\pi}(E')$ acquires a nonzero imaginary part
below the (in--medium) two--pion threshold all the way down to
$E'=0$. This leads to an additional contribution to the
$E'$--integral in eqs.~(19),~(28), accounting for the almost entire
difference in comparison to scheme 'a'.

\noindent
We end this chapter by testing our numerical results with respect to
thermodynamic consistency. For that we recalculate the total
pressure--density by integrating the entropy--density according
to eq.~(49), which amounts to an implicit check of the stationarity
condition, eq.~(44). As can be seen from fig.~2 there
is excellent agreement of the such calculated $p_\pi(T)$
with the 'direct' evaluation when neglecting the
pion width (scheme 'a'). When including a finite $Im \Sigma_\pi$
(scheme 'b') the values for pressure--density obtained from eq.~(49)
are slightly suppressed as compared to the 'direct' calculation,
causing a loss of about half the enhancement over scheme 'a'.

\section{Dropping Rho Meson Masses}
In the calculations of the previous chapter no medium effects were
considered in the interaction kernel of the $\pi\pi$ scattering
equation. However, since our $\pi\pi$ model is based on explicit
(vector) meson exchange, one may expect an impact on the
pseudopotentials due to modifications in the propagator of the
exchanged mesons when exposed to finite temperature.
The most prominent example for such a modification is the dropping
of the vector meson masses or
'Brown--Rho Scaling'~\cite{BrRo,Hats}. Adami and Brown suggested
the temperature dependence of the $\rho$ meson mass to be~\cite{AdBr}
\beq
m_\rho(T)\approx m_\rho(0) \ \biggl (\frac{<\bar qq>_T}
{<\bar qq>_0}\biggr )^{1/3} \ ,
\eeq
where $<\bar qq>_T$ denotes the quark condensate at finite
temperature.
The $T$ dependence was taken as
\beq
<\bar qq>_T=<\bar qq>_0 \sqrt{1-(T/T_c^\chi)^2} \ .
\eeq
In ref.~\cite{Rdth} such a decrease of the $\rho$ mass was implemented
in a selfconsistent pion gas calculation within the J\"ulich model
as described in sect.~3.2.. Already well below $T_c^\chi$ a large
accumulation of strength was found in the scalar--isoscalar channel
of the $\pi\pi$ scattering amplitude close to the two--pion threshold.
It is due to the strong enhancement of the attractive t--channel
$\rho$ exchange in this channel.  However, it has been shown
recently~\cite{ARCSW} that the implementation of chiral constraints
into the $\pi\pi$ interaction is crucial to reliably calculate
in--medium $\pi\pi$ correlations in the vicinity of the
threshold. On the other hand, sufficiently above the
two--pion threshold ($E_{cms}\ge 400 MeV$ or so)
chiral constraints rapidly cease to
have significant impact on in--medium $\pi\pi$ amplitudes. This
justifies the use of the non--chirally symmetric J\"ulich model
in the previous chapter, where no dropping vector meson masses are
taken into account and thus only minor threshold effects in the
$\pi\pi$ amplitude are observed~\cite{RW1}.  \\
In this chapter we will employ a chirally improved version of the
J\"ulich $\pi\pi$ interaction~\cite{DRW,RW3}. It is supplemented
with $\pi\pi$ contact interactions as required from the
gauged nonlinear $\sigma$ model~\cite{W68}. To ensure the correct
chiral limit for the s--wave scattering lengths a modified
off--shell prescription for the pseudopotentials has been chosen
when iterating the scattering eq.~(14).  \\
Due to the additional interaction terms in the Lagrangian of the
chirally improved J\"ulich model there arise further contributions
to the thermodynamic potential $\Omega_\pi(T)$. They are
straightforwardly incorporated in the quasiparticle and in the
lowest order contributions, $\Omega_{\pi}^Q(T)$ and
$\Omega_{\pi\pi}^{(1)}(T)$, respectively. For
$\Omega_{\pi\pi}^{(n)}(T)$ we take into account all
contributions up to second order, namely from
\begin{itemize}
\item[--]
second--order $\rho$ exchange as given by eq.~(19) for $n$=2
with
\beq
\frac{1}{8} Im M_{\pi\pi}^{(2)}=\frac{1}{8} V_{\rho ex} \
Im G_{\pi\pi} \ V_{\rho ex} \ ,
\eeq
where we explicitly indicated the degeneracy factor 1/4$n$;
\item[--]
the second--order contact interactions also given by eq.~(19)
for $n$=2, but with the degeneracy factor $\frac{1}{8}$
replaced by $\frac{1}{4}$ due to the absence of exchange
diagrams (compare the remarks following eq.~(4)), {\ie}
\beq
\frac{1}{4} Im M_{\pi\pi}^{(2)}=\frac{1}{4} V_{cont} \
Im G_{\pi\pi} \ V_{cont} \ ;
\eeq
\item[--]
the first--order $\rho$ exchange plus first--order contact
interaction; here the appropriate degeneracy
factor is $\frac{1}{6}$:
\beq
\frac{1}{6} Im M_{\pi\pi}^{(2)}=\frac{1}{6} V_{cont} \
Im G_{\pi\pi} \ V_{\rho ex} \ .
\eeq
\end{itemize}
Numerically it turns out that higher--order contributions in
$\rho$ exchange or contact interactions are again negligible. \\
When recalculating the EOS without any dropping of the $\rho$
mass the results employing the chirally improved J\"ulich
model differ by around 5\% from those presented in sect.~3.3,
confirming
the notion that chiral symmetry has little impact on the $\pi\pi$
interaction as long as strong threshold effects are absent.

\noindent
Let us now turn to the scenario with 'Brown--Rho Scaling'
included. For simplicity we will neglect the coupling to the
$K\bar K$ channel, which was shown to have a very small effect
in the previous chapter. Relevant for our model are the decrease
of the (physical) $\rho$ mass (eq.~(53)) and the corresponding
reduction of the pion decay constant,
\beq
\frac{f_\pi(T)}{f_\pi(0)}=\frac{m_\rho(T)}{m_\rho(0)} \ .
\eeq
The latter enters the $\pi\pi$ contact interactions, whereas the
physical mass $m_\rho$ is used in the t--channel $\rho$
exchange propagator, eq.~(33). For given temperature the bare
$\rho$ mass $m_\rho^{(0)}$ used in the s--channel $\rho$ pole
propagator, eq.~(23), is adjusted such that the fully renormalized
$M_{\pi\pi}^{11}$--amplitude in free space acquires it's maximum
({\ie}$\rho$ resonance peak) at the corresponding value of
$m_\rho(T)$ from eqs.~(53),~(54). For fixed $m_\rho(T)$, $f_\pi(T)$
and $m_\rho^{(0)}(T)$ we then perform selfconsistent Brueckner
calculations for $\Sigma_\pi$ and $M_{\pi\pi}$ at several $T$--values
along exactly the same lines as described in sect.~3.2. \\
The resulting EOS for the hot interacting pion gas is displayed in
fig.~3, where we have chosen the critical
temperature of chiral symmetry
restoration to be $T_c^\chi$=170~MeV. For T$\geq$100~MeV we observe
a strong increase of the thermodynamic state variables as compared
to the calculations without inclusion of the Brown--Rho Scaling.
Most of the effect is of simple kinematic origin, as can be seen
by comparison to the free $\pi$--$\rho$ gas including the decrease
of $m_\rho$. However, for temperatures below $T\approx 150$~MeV
the $\pi\pi$ interactions still generate an enhancement over the free
$\pi$--$\rho$ case of up to 15\%, which is
quite similar to what we found in the pure Brueckner calculations.
Just below $T_c^\chi$ the free $\pi$--$\rho$ gas values
for $p_\pi(T)$, $s_\pi(T)$ and $\epsilon_\pi(T)$ are close to
the ones of the interacting pion gas. This is due to the
fact that the $\rho$ polegraph starts acquiring appreciable
renormalization contributions especially from the strongly attractive
t--channel exchange of a $\rho$ meson of mass $m_\rho(T)$. However,
in our approximation scheme for $\Omega_{\pi\pi}(T)$, this kind of
contributions is not accounted for (compare sect.~2.2.1), even though
they are certainly not negligible anymore
in the vicinity of $T_c^\chi$.

\section{Summary and Conclusions}
Based on the finite--temperature Greens--function formalism we have
presented an analysis of the equation of state of a hot interacting
gas of pions at zero chemical potential. \\
Starting from a realistic $\pi\pi$ meson--exchange model
capable of describing the vacuum scattering data over a
broad range of energies we have calculated the in--medium $\pi\pi$
scattering
amplitude and the single--pion selfenergy. The resulting
selfconsistency
problem has been solved by numerical iteration within
the quasiparticle approximation (Brueckner scheme)~\cite{RW1}. \\
We then proceeded to evaluate the thermodynamic potential
$\Omega_\pi(T)$:
\begin{itemize}
\item[--] the single--pion selfenergy was used for calculating
         the quasiparticle contribution $\Omega_\pi^Q(T)$; the latter
         turned out to be very close to the values for free pions
         due to a cancellation between the mean field term and the
         free quasi--pion term (only when including a finite width for
         the pions a $\approx$5\% enhancement was found);
\item[--] for the interaction part of the thermodynamic potential,
         $\Omega_{\pi\pi}(T)$, we have taken into account the
         $f_0(980)$--,
         $\rho(770)$-- and $f_2(1270)$--s--channel pole graphs to all
         orders as well as the $\rho$ t--channel exchange up to third
         order; higher orders as well as interference terms between
         s-- and t--channel graphs have been estimated to be small
         and were neglected; the main contribution stems
         from the $\rho$
         pole graphs, their magnitude being comparable to what one
         expects from an admixture of free $\rho$ mesons.
\end{itemize}
Special attention has been paid to constraints from thermodynamic
consistency, and it was shown that our results satisfactorily fulfill
these constraints. \\
In total we have found a 10--15\% enhancement of
$\Omega_\pi(T)$ compared
to a free gas of $\pi$ and $\rho$ mesons in the temperature range
100~MeV$\leq$~T~$\leq$ 150~MeV. Pressure--, entropy-- and
energy--density,
which were extracted from $\Omega_\pi(T)$ by means of standard
thermodynamic relations, show a very similar behavior. This is in
qualitative agreement with the results of Welke et al.~\cite{WVPr},
who employed empirical vacuum s-- and p--wave $\pi\pi$ phase shifts
within a relativistic virial expansion. Our results are, however, at
qualitative variance with the findings of Bunatian and
K\"ampfer~\cite{BuKa}, who also used the
finite temperature Greens--function approach
but employed the Weinberg Lagrangian. The latter
is known to account only for the low--energy s--wave $\pi\pi$
interaction, resulting in a net repulsion when performing the
isospin weighted sum for the near--threshold $\pi\pi$ amplitude. As
a consequence, their thermodynamic state variables show an overall
{\it decrease} with temperature as compared to the free pion gas.  \\
Our conclusion from this is that the higher energy range (up to
CMS energies close to 1~GeV) as well as higher partial waves
(especially the resonant p--wave) in the $\pi\pi$ interaction
are important for a reliable description of thermodynamic
properties of an interacting gas of pions. \\
We furthermore investigated the impact of a dropping $\rho$ mass
according to  'Brown--Rho Scaling'. Using a chirally improved
J\"ulich model including contact interactions, arising from the
gauged Weinberg Lagrangian, we find a similar behavior as before:
even though the thermodynamic state
variables of the interacting pion gas now exhibit a considerable
increase near $T_c^\chi$, the enhancement over the free
$\pi$--$\rho$ gas values (dropping $\rho$ mass included) is not
more than 10--20\%. Very close to the critical temperature, where the
$\rho$ mass rapidly drops to zero, the interacting pion gas values
are close to the ones of the free $\pi$--$\rho$ gas. This feature,
however, is due to the breakdown of our approximations; in particular,
interference terms between $\rho$ s-- and t--channel graphs become
far from negligible.

the strong
gauge 'data'.
('external')
however,

\vspace{2cm}
\bce
{\bf Acknowledgement}
\ece
\vspace{0.5cm}

\noindent
We thank  G.G. Bunatian, D. Blaschke and  J. W. Durso
for useful discussions.
One of us (R.R.) acknowledges financial support by Deutscher
Akademischer Austauschdienst (DAAD) under program HSPII/AUFE.
This work was supported in part by a grant from the National Science
Foundation, NSF-PHY-94-21309.
\vfill\eject

\vfill\eject

\pagebreak

%
\bce
{\bf Tables}
\ece
\vskip 1.0cm
\begin{table}[h]
\begin{tabular}{|c|ccc|}
\hline
T [MeV] & $p_{\pi\pi\rho}/p_{\pi}^{free}$ &
$s_{\pi\pi\rho}/s_{\pi}^{free}$ &
$\epsilon_{\pi\pi\rho}/\epsilon_{\pi}^{free}$  \\
\hline
150 &  26\%  & 41\%  & 45\%  \\
200 &  43\%  & 56\%  & 60\%  \\
\hline
\end{tabular}
\caption{Contributions from the $\rho$ pole graph skeleton diagrams
to the thermodynamic state variables in an interacting hot pion gas
within the Brueckner calculations neglecting any width of the pions.}
\end{table}

\vspace{0.5cm}

\begin{table}[h]
\begin{tabular}{|c|ccc|}
\hline
T [MeV] & $p_{\pi\pi\rho ex}^{(2)}/p_{\pi}^{free}$ &
$s_{\pi\pi\rho ex}^{(2)}/s_{\pi}^{free}$ &
$\epsilon_{\pi\pi\rho ex}^{(2)}/\epsilon_{\pi}^{free}$  \\
\hline
150 &  7.0\%  & 8.5\%   & 8.9\%   \\
200 &  9.8\%  & 11.2\%  & 11.7\%  \\
\hline
\end{tabular}
\caption{Contributions from the 2.order $\rho$ t--channel exchange
skeleton diagrams to the thermodynamic state variables of an
interacting hot pion gas within
the Brueckner calculations, neglecting any width of the pions.}
\end{table}

\vspace{0.5cm}

\begin{table}[h]
\begin{tabular}{|c|cccc|}
\hline
        & $p_{\pi\pi\rho ex}^{(2)}/p_{\pi}^{free}$ &
$p_{\pi\pi\rho}/p_{\pi}^{free}$ & $p_\pi^Q/p_{\pi}^{free}$
& $p_\pi^{tot}/p_{\pi}^{free}$   \\
\hline
$Im\Sigma_\pi\equiv 0$ &  9.8\%  & 43\%   & 99.8\%  & 158.4\% \\
$Im\Sigma_\pi<0$  &  13.3\%  & 46.6\%  & 107.6\% & 173.7\% \\
\hline
\end{tabular}
\caption{Various contributions to the pressure density of an
interacting hot pion gas at $T=200$~MeV within the Brueckner
calculations when neglecting (upper line) and
including (lower line) a finite pion width.}
\end{table}
\vfill\eject

\newpage

\bce
{\bf Figure Captions}
\ece
\vskip 1.0cm
\begin{itemize}
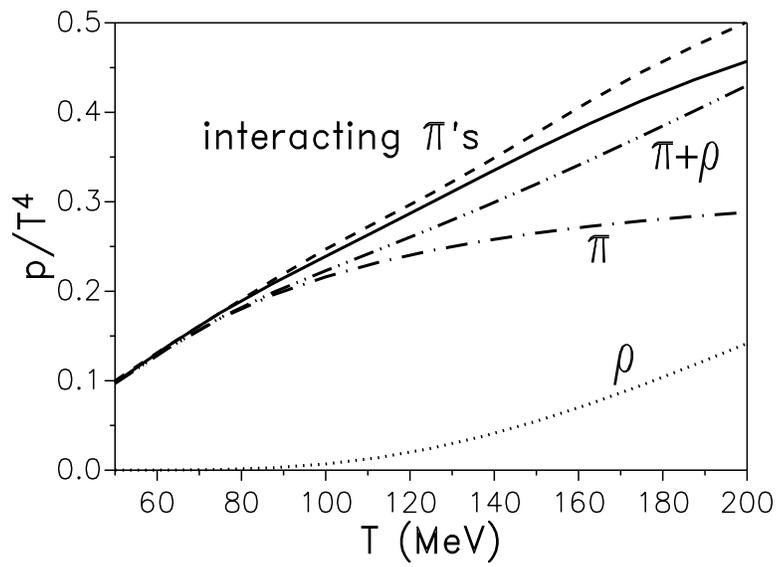
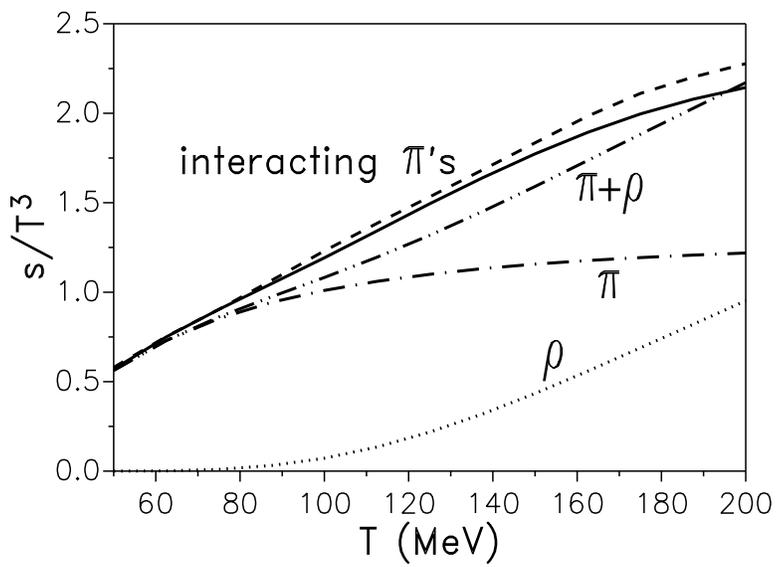
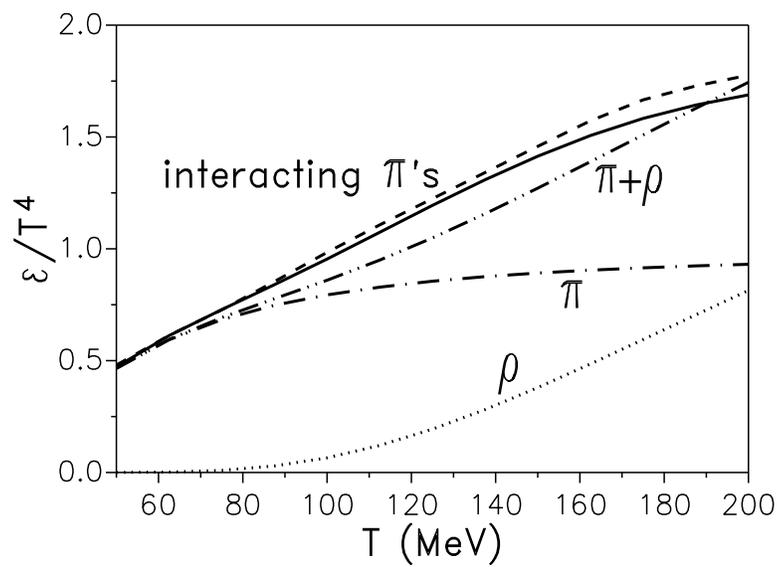
\item[{\bf Fig.~1}:]
Thermodynamic state variables (scaled by temperature to
dimensionless units) of an interacting hot pion gas within
the Brueckner scheme  with no medium modifications applied to
the two--body interaction potentials; \\
upper panel: pressure--density; \\
middle panel: entropy--density; \\
lower panel: energy--density; \\
full lines: interacting pion gas when neglecting the pion width;
dashed lines: interacting pion gas with inclusion of a finite
pion width; dotted lines: free $\rho$ gas; dashed--dotted lines:
free pion gas; dashed--double--dotted lines: free $\pi$--$\rho$ gas.

\vskip 0.5cm

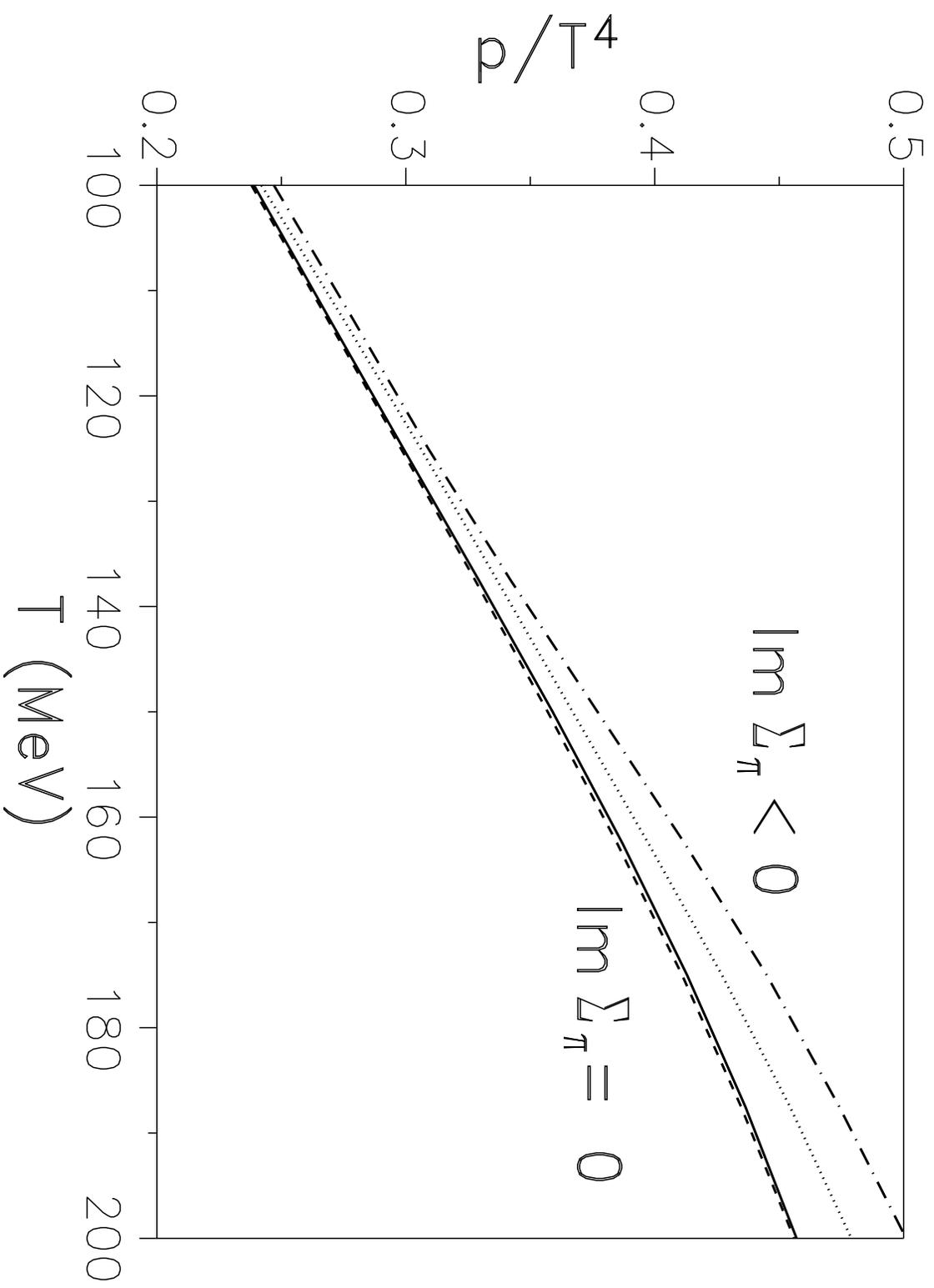
\item[{\bf Fig.~2}:]
Comparison of two different ways of calculating the pressure--density
of an interacting pion gas within the Brueckner scheme;
the full line ($Im \Sigma_\pi$ neglected) and the dashed--dotted
line ($Im \Sigma_\pi$ included) correspond to the 'direct'
calculations (the sum of eqs.~(12),~(19),~(28) and~(39)/(40)),
whereas the dashed line ($Im \Sigma_\pi$ neglected) and the
dotted line ($Im \Sigma_\pi$ included) are obtained from integrating
the entropy--density according to eq.~(49).

\vskip 0.5cm

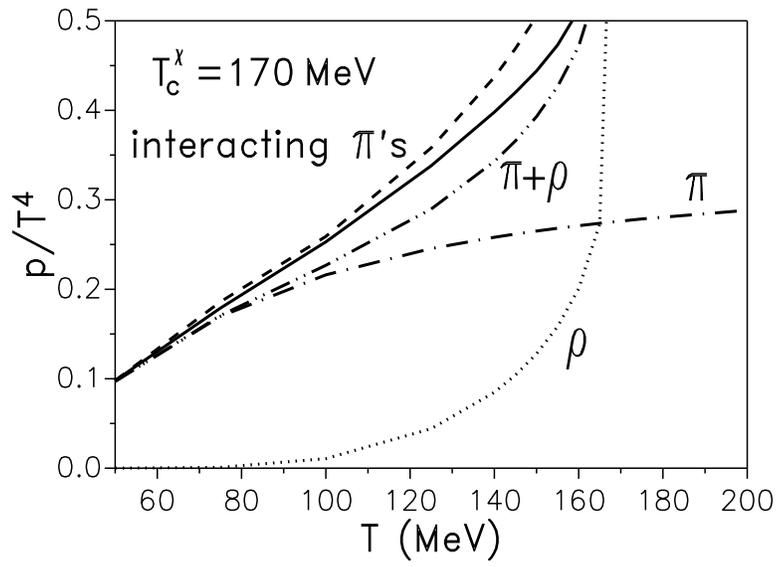
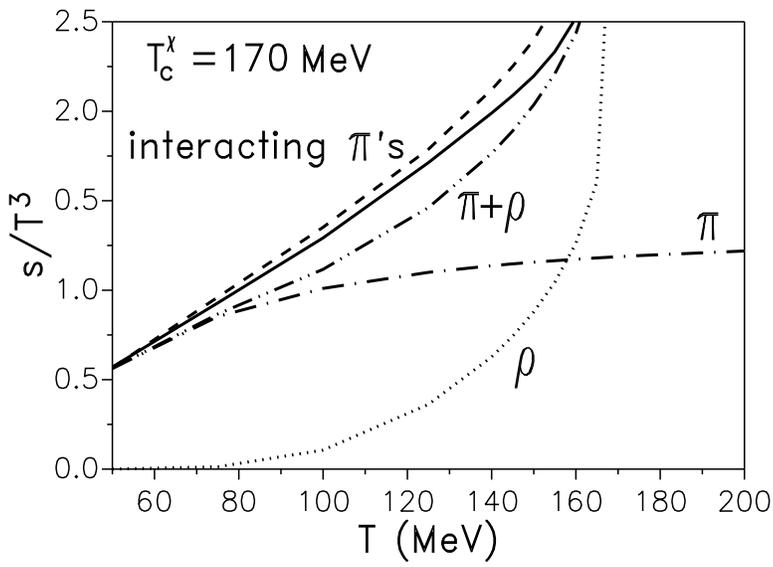
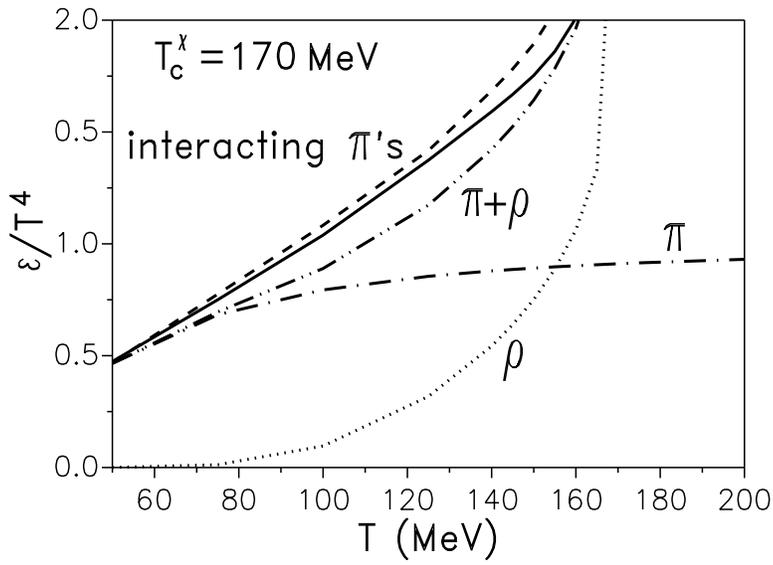
\item[{\bf Fig.~3}:]
Thermodynamic state variables (scaled by temperature to
dimensionless units) of an interacting hot pion gas within
the Brueckner scheme plus additional medium modifications
of the two--body pseudopotentials in form of dropping
$m_\rho(T)$ and $f_\pi(T)$ ; here, a chirally improved version
of the J\"ulich $\pi\pi$ interaction has been employed; \\
upper panel: pressure--density; \\
middle panel: entropy--density; \\
lower panel: energy--density; \\
line identification as in fig.~1.



\end{itemize}

\end{document}